\documentclass[aps,prl,reprint,amsmath,amssymb,showpacs]{revtex4-1}
\usepackage{times}
\usepackage{graphicx}
\usepackage{subfigure}
\usepackage{dcolumn}
\usepackage{bm}
\def \bea{\begin{eqnarray}}
\def \eea{\end{eqnarray}}

\begin{document}


\title{Jamming Percolation in Three Dimensions}
\author{Antina Ghosh}
\author{Eial Teomy}
\author{Yair Shokef}
\email{shokef@tau.ac.il}
\affiliation{School of Mechanical Engineering, Tel-Aviv University, Tel-Aviv 69978, Israel}

\begin{abstract}

We introduce a three-dimensional model for jamming and glasses, and prove that the fraction of frozen particles is discontinuous at the directed-percolation critical density. In agreement with the accepted scenario for jamming- and glass-transitions, this is a mixed-order transition; the discontinuity is accompanied by diverging length- and time-scales. Because one-dimensional directed-percolation paths comprise the backbone of frozen particles, the unfrozen rattlers may use the third dimension to travel between their cages. Thus the dynamics are diffusive on long-times even above the critical density for jamming.

\end{abstract}

\pacs{64.70.Q-,64.60.ah,05.50.+q}


\maketitle

\section{Introduction}

Glassiness and jamming in grains and colloids entails mechanical solid-like behavior accompanied by divergent relaxation times, apparently without spatial order~\cite{vanHecke_Review_2010,Liu_Nagel_Review_2010,Biroli_Garrahan_Perspective_2013}. A rather simple way to theoretically study such non-equilibrium transitions is by lattice models in which disorder and slow (or nonexistent) dynamics arise naturally from the underlying static or dynamic local rules of the models~\cite{Biroli_Mezard_Lattice_Glass,PicaCiamarra_Lattice_Glass,Eisenberg_Baram,Rotman_Eisenberg}. Thus, a promising approach for describing the glass and jamming transitions has evolved in recent years around \emph{kinetically-constrained models}~\cite{Ritort_Sollich,Garrahan_review}. The first such model due to Fredrickson and Andersen~(FA)~\cite{FA} considers non-interacting spins on an ordered lattice such that each spin can flip only if at least some number $m$ of its neighbors are up. Kob and Andersen~(KA) later suggested the corresponding lattice-gas model~\cite{KA}. Although these two models were thought to exhibit a glass transition, it was recently shown that they jam only due to finite-size effects~\cite{Holroyd,Gravner,TBF_2004,Balogh,Teomy}. Namely, for any temperature (FA) or density (KA) there is a finite length-scale, such that in systems larger than this scale, all particles eventually move. Subsequently, two-dimensional~(2D) jamming-percolation models were introduced and proven to jam at finite temperature (spin models) or density (lattice gases) even in the thermodynamic limit~\cite{TBF_2006,Jeng_Schwarz_comment,TBF_reply,spiral_EPJB,spiral_JSP,JengSchwarz}. In such models, the kinetic rules defining which particles are constrained from flipping (or moving) depend not only on the number of up (or occupied) neighbors but also on their relative positions. Consistently with the glass and jamming scenarios, these models exhibit a phase-transition of mixed nature~\cite{Kirkpatrick,Durian_1995,Yin,Biroli_Bouchaud_EPL2004,Sellitto_EPL2005,Schwarz_Liu_Chayes,Majmudar}; the fraction of frozen particles jumps discontinuously, yet there are critical scaling and diverging length- and time-scales.

Glass- or jamming-transitions result from the fact that particles are \emph{blocked} by their neighbors, which in turn are blocked by their neighbors, and so on, hence forming large clusters of particles that cannot move. In jamming-percolation models, the blocked particles form similar clusters, which above the critical density become infinite and span the entire system, thus a finite fraction of the system is completely \emph{frozen}. In spite of the qualitative differences between percolation in two and three dimension, most work so far was on 2D models, and it is not clear to what extent the results obtained there would be relevant for actual 3D systems.

\begin{figure}
\includegraphics[width=\columnwidth, bb= 92 297 540 510, clip=true]{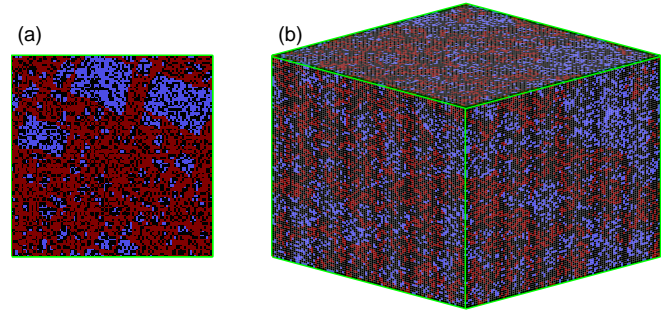}
\caption{Above the critical density, in the 2D spiral model (a) movable particles (blue) are confined by permanently frozen (red) particles, while in our 3D model (b), they can use the third dimension to travel within the sponge-like, percolating frozen structure. 
Representative configurations of linear system size L=100 with particle densities $\rho=0.64(2D),0.35(3D)$.}
\label{fig:face}
\end{figure}

In this letter we propose a 3D kinetically-constrained model, for which we prove there is a mixed-order jamming transition at finite density. Our numerical investigation supports the theoretical predictions we give here that the static properties of this model are qualitatively similar to those of the 2D spiral model \cite{spiral_EPJB,spiral_JSP}, whereas its dynamics are qualitatively different. In both models, jamming occurs because 1D strings of blocked particles span the system. In 2D, these strings of frozen particles confine the mobile particles in compact domains, see fig.~\ref{fig:face}a, hence these cannot diffuse on long time scales. Our 3D model is also jammed by 1D system-spanning frozen strings. However, the mobile particles in this case can use the third dimension to bypass the frozen strings, see fig.~\ref{fig:face}b, and thus they exhibit long-time diffusive behavior even above the critical density for jamming. Some attempts have been previously made to extend several force-balance-percolation models to 3D~\cite{JengSchwarz}, but to our knowledge, we are the first to provide exact results, and to study the dynamics of such models in three dimension.

\section{Model}

\begin{figure}
\includegraphics[width=\columnwidth, bb=24 349 620 504, clip=true]{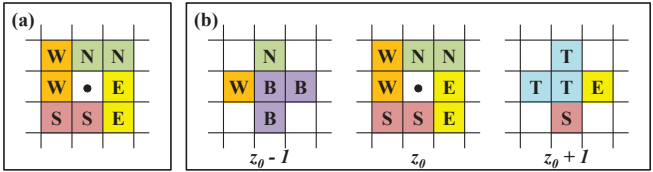}
\caption{(a) The 2D spiral model is defined by the sets West(W), East(E), South(S), North(N). (b) Our 3D model is defined by adding sets Bottom(B), Top(T), and by including a third particle in each of the Spiral model's sets. Slices from three consecutive layers of the cubic lattice are shown side by side.}
\label{fig:rules}
\end{figure}

The 2D spiral model~\cite{spiral_EPJB,spiral_JSP,Corberi,Shokef_Liu} was defined on the square lattice by dividing each site's 4 nearest- and 4 next-nearest-neighbors into four pairs of adjacent sites as depicted in fig.~\ref{fig:rules}a. Here, the central site (\textbullet) is unblocked if its ((W {\it or} E) {\it and} (S {\it or} N)) pairs do not contain any particles. Inspired by the 2D spiral model, we define our 3D model on the cubic lattice, and divide each site's 6 nearest- and 12 next-nearest-neighbors into 6 sets, each set consisting of one nearest neighbor and two next-nearest neighbors as shown in fig~\ref{fig:rules}b. We define the central site (\textbullet) as unblocked if its ((W {\it or} E) {\it and} (S {\it or} N) {\it and} (B {\it or} T)) sets are completely empty. At each time step of the dynamics of both models a particle is randomly chosen and moved to one of its randomly selected nearest-neighbor sites only if the target site is vacant and the particle is unblocked both before and after the move. Here we focus on these lattice-gas dynamics, in which there is sense to motion and hence to diffusion, yet our static results are equally valid for the Ising variant of this model, in which individual unblocked sites can flip their state at some temperature-dependent rates.

\begin{figure}
\includegraphics[width=\columnwidth]{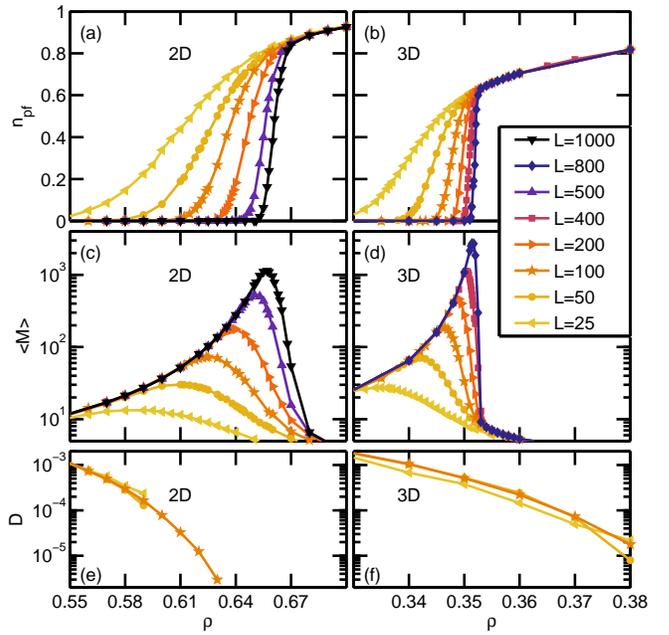}
\caption{Fraction of permanently frozen particles (a,b), Mean culling time (c,d), and Diffusion coefficient (e,f) vs. particle density, for the 2D spiral model (a,c,e) and for our 3D model (b,d,f). Legend: linear dimension $L$ of the lattices.}
\label{fig:npf}
\end{figure}

Our 3D kinetic constraint is clearly more restrictive than that of the 2D spiral model, since in each of the three orthogonal planes passing through a given site in the cubic lattice we have the 2D spiral rules. Thus for $\rho>\rho_c^{2D}$ there is a finite fraction of permanently-frozen particles, with $\rho_c^{2D}\approx0.71$ the critical density of 2D directed percolation~(DP), at which the 2D spiral model jams~\cite{spiral_EPJB,spiral_JSP}. Before proving that our 3D model undergoes a discontinuous transition at the critical density $\rho_c^{3D}\approx0.43$~\cite{DeBell,Adler} of 3D~DP, we present in fig.~\ref{fig:npf},a-b numerical results for the fraction of permanently frozen particles vs. density. These indicate a jamming transition at $\rho_c \approx 0.4$ in our 3D model and at $\rho_c \approx 0.7$ in the 2D spiral model.

\section{Proof} 

\begin{figure}
\includegraphics[width=\columnwidth, bb=344 244 692 498, clip=true]{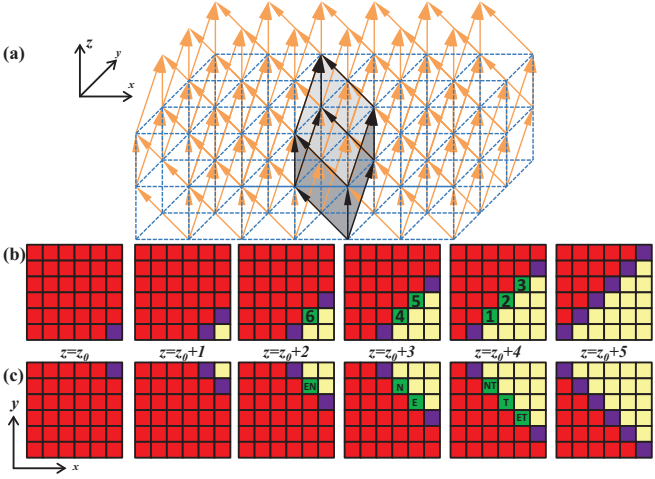}
\caption{(a) Directed cubic lattice formed by drawing arrows from each site to its three T neighbors. Unit cell highlighted in black. (b-c) If ${\cal T}^{\vec{a}}_5$ (yellow) and the diagonals of ${\cal T}^{\vec{a}}_6$ (purple) are empty, then for $\vec{a}=(-+-)$ (b) the remaining sites (green) of ${\cal T}^{\vec{a}}_6$ may be emptied by the E-S-T sets in the order indicated by the numbers. For $\vec{a}=(---)$ (c) the remaining sites have only one or two of their E, N and T sets empty as indicated in the figure, thus they are not necessarily emptiable.}
\label{fig:tetrahedron}
\end{figure}

We will prove that in our model the number of frozen particles is macroscopic for $\rho>\rho_c^{3D}$, and that for $\rho <\rho_c^{3D}$ in the thermodynamic limit there are no frozen particles. Together this implies that the fraction of frozen particles jumps discontinuously at $\rho_c^{3D}$ from zero to a finite value. Our proof extends to three dimensions the corresponding theoretical work regarding the 2D spiral model~\cite{spiral_EPJB,spiral_JSP}. Yet, as will become evident below, our extension is not as straightforward as expected. We start by considering the directed cubic lattice formed by drawing arrows from each lattice site to its three T neighbors, see fig~\ref{fig:tetrahedron}a. This maps the kinetic constraint to 3D DP, thus for $\rho>\rho_c^{3D}$ there is an infinite sequence of frozen particles which lie along a B-T path, and a finite fraction of frozen particles even in the thermodynamic limit.

To prove that for $\rho<\rho_c^{3D}$ there are no frozen particles, as opposed to the proof for the 2D spiral model, here we will demonstrate that with finite probability a small unblocked region can expand to unblock only six-eights of the system, and not all of it. Since in an infinite system there are an infinite number of such finite initially unblocked regions, the entire system can thus be unblocked. To simplify the presentation, we consider \emph{culling} dynamics, in which unblocked particles are \emph{removed} rather than \emph{moved} to neighboring sites. This immediately proves the transition for Ising spin-flip dynamics, and we expect that the same results for the aforementioned lattice-gas dynamics should follow.

Around any lattice site $\vec{r}_0=(x_0,y_0,z_0)$, for any direction $\vec{a}=(a_x,a_y,a_z)$ from it to one of its third order neighbors, with $a_{i}=\pm1$, and for any integer distance $\ell$, we define the \emph{trirectangular tetrahedron} ${\cal T}^{\vec{a}}_{\ell}(\vec{r}_0)$ as all sites $\vec{r}=(x,y,z)$ for which $(x-x_0)a_x\ge 0$, $(y-y_0)a_y\ge 0$, $(z-z_0)a_z\ge 0$, and $|x-x_0|+|y-y_0|+|z-z_0|<\ell$. The \emph{legs} of this tetrahedron are its edges which lie along the $\hat{x}$, $\hat{y}$ and $\hat{z}$ directions, and its \emph{diagonals} are its other three edges. Now assume that for a given $\ell>3$ all sites in ${\cal T}^{\vec{a}}_{\ell-1}$ are empty, and that the three diagonals of ${\cal T}^{\vec{a}}_{\ell}$ are empty, and ask whether all sites in the larger tetrahedron ${\cal T}^{\vec{a}}_{\ell}$ may also be emptied. Fig.~\ref{fig:tetrahedron},b-c shows that for $\vec{a}=(-+-)$ this is possible, yet for $\vec{a}=(---)$ it is not. One can verify that such expansion is possible for all $\vec{a}$ vectors except for $\vec{a}=\vec{1}\equiv(1,1,1)$ and $\vec{a}=-\vec{1}$. We thus define the almost-octahedron ${\cal O}_{\ell}(\vec{r}_0)$ as the union of the six tetrahedra ${\cal T}^{\vec{a}}_{\ell}(\vec{r}_0+\frac{\vec{a}+\vec{1}}{2})$ for all $\vec{a}$ vectors except for $\vec{a}=\pm\vec{1}$. Applying the above result for each tetrahedron, we deduce that ${\cal O}_{\ell}$ may be emptied if ${\cal O}_{\ell-1}$ is empty and all sites of the 18 diagonals of ${\cal O}_{\ell}$ can be emptied.

Now, for a site in location $1 \leq s \leq \ell$ along a diagonal of ${\cal O}_{\ell}$ to be emptiable, there should not be a directed path from this site to any of the sides of the cube confining the tetrahedron to which it belongs. The lengths of these three paths are $s$, $\ell-s+1$ and $\ell$. Thus, the probabilities of having them are $e^{-s/\xi}$, $e^{-(\ell-s+1)/\xi}$ and $e^{-\ell/\xi}$, respectively (see Appendix A), with $\xi$ the DP correlation length, which diverges at $\rho_c^{3D}$. Due to the positive correlations between probabilities of different sites on the diagonal to be emptiable, the probability that ${\cal O}_{\ell-1}$ may be expanded to ${\cal O}_{\ell}$ is bounded by (see Appendix B):
\begin{eqnarray}
P_{\ell} &\ge& 18 \prod_{s=1}^{\ell} \left(1-e^{-s/\xi}\right)\left(1-e^{-(\ell-s+1)/\xi}\right) \left(1-e^{-\ell/\xi}\right) \nonumber\\ &\geq& C \exp\left(-\ell e^{-\ell/\xi}\right) \label{eq:one}
\end{eqnarray} 
with $C>0$. The probability to expand ${\cal O}_{\ell}$ to infinity is thus 
\begin{eqnarray}
\prod^{\infty}_{\ell=1}P_{\ell} &\geq& \prod^{\infty}_{\ell=1}C\exp\left(-\ell e^{-\ell/\xi}\right) = C\exp\left(-\sum^{\infty}_{\ell=1}\ell e^{-\ell/\xi}\right) \nonumber\\ &=& C\exp\left[-e^{1/\xi} / \left(1-e^{1/\xi}\right)^{2}\right]>0, \label{eq:two}
\end{eqnarray}
where the last inequality follows from $\xi$ being finite. Thus for $\rho<\rho_c^{3D}$, with finite probability a small region may be expanded to an infinite almost-octahedron, and empty six-eights of the system. Now, any site $\vec{r}=(x,y,z)$ in the lattice may be emptied if there is an almost-octahedron centered at any other site $\vec{r}_0=(x_0,y_0,z_0)$ except for those with $sign(x-x_0)=sign(y-y_0)=sign(z-z_0)$. As there is an infinite number of such $\vec{r}_0$ sites, any $\vec{r}$ can be emptied, and for $\rho<\rho_{c}^{3D}$ there are no blocked particles in the infinite-system-size limit.

\begin{figure}
\includegraphics[width=\columnwidth, bb=8 13 420 174, clip=true]{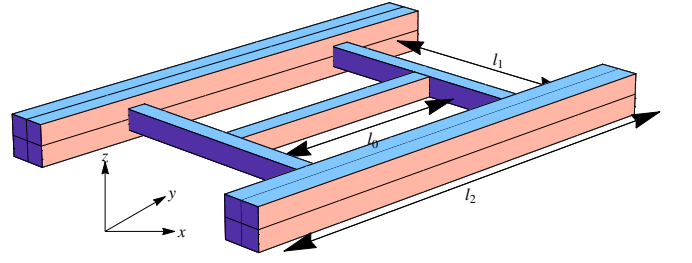}
\caption{First three sets of rhombi blocking each other.}
\label{fig:rhomboids}
\end{figure}

To conclude the proof, we will now show that the fraction of frozen particles is discontinuous, namely at $\rho_c^{3D}$ each site has a finite probability of being frozen. Extending the proof for the 2D spiral model, we start with a tilted rhombus oriented along the S-N direction of some size $l_0\times kl_0\times kl_0$, with $k=\frac{1}{12}$. Starting from the two far edges of this rhombus we construct the series shown in fig.~\ref{fig:rhomboids} of pairs of tilted rhombi of size $l_i\times kl_i\times kl_i$ that alternate in their long direction between the S-N and W-E directions, such that the far ends of each rhombus touch the next two rhombi, and with $l_{1}=l_{0}$ and $l_i=2l_{i-2}$. The value $k=\frac{1}{12}$ is chosen so that different rhombi will not intersect. The original site is frozen if all these rhombi have DP paths along their long directions and each path is connected to the previous path. For any density and for any length $l_{0}$, there is some finite probability $p$ that the first rhombus contains a DP path of length $l_{0}$ in the S-N direction, and that this path continues until the edges of the two rhombi adjacent to the first one, so that its total length in the S-N direction is $\left(1+2k\right)l_{0}$. Now, consider each of the adjacent rhombi and divide it into $l^{2\left(1-1/z\right)}_{1}$ channels, each of size $l_{1}\times kl^{1/z}_{1}\times kl^{1/z}_{1}$, with $z>1$ the DP exponent relating length and width of DP paths. Namely, typical DP clusters of parallel length $l$ have typical transverse length $l^{1/z}$~\cite{DeBell,Adler,DP_exponents}. By this construction, for $\rho>\rho_c^{3D}$, each such channel contains a DP path with some finite probability $q$, which is independent of $l_{1}$. The path spanning the first rhombus passes through $l^{1-1/z}_{1}$ such channels, and so the probability that at least one of these channels will have a DP path spanning it is bounded from below by $1-q^{l^{1-1/z}_{1}}$. This process can be continued indefinitely since the probability that the original site is frozen is bounded from below by $p\prod^{\infty}_{i}\left(1-q^{l^{1-1/z}_{i}}\right)^{2}$, which is finite.

\section{Critical Scaling} 

\begin{figure}
\includegraphics[width=\columnwidth]{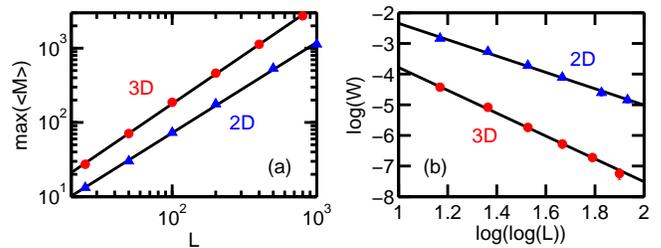}
\caption{(a) Maximal culling time vs. system size fits ${\rm max} (\langle M \rangle) \propto L^\alpha$ with $\alpha=1.22(2D), 1.33(3D)$. (b) Transition width fits $W\propto (\ln L)^{-1/\mu}$ with $\mu=0.38(2D), 0.27(3D)$. }
\label{fig:exponents}
\end{figure}

We have shown that the fraction of frozen particles is discontinuous at $\rho_c^{3D}$. As discussed above, the phase transition we observe in our 3D model also exhibits some features of continuous transitions. To probe time- and length-scales we run the following culling dynamics both for our 3D model and for the 2D spiral model. We start with a randomly occupied lattice at a given density. We then check the kinetic constraint on all sites and identify the mobile particles. In a single culling step all mobile particles are removed, and subsequently other particles blocked by them may become mobile. These in turn are removed in the following step, and this process is continued iteratively until either the lattice is empty or no more particles can be removed. We obtain a \emph{time scale} from the average number of culling iterations $\langle M\rangle$ needed to remove all movable particles. This is related to the time it would take a particle, which is not permanently frozen to eventually relax under the system's actual dynamics. Clearly, $\langle M \rangle$ is not directly proportional to the structural relaxation time or to the persistence time in the system, but we expect to find strong correlations between these two measures. Near $\rho_c$ not many particles can initially move, and one has to cull many layers of particles until the frozen core is reached. Thus, $\langle M\rangle$ peaks at $\rho_c$, as shown in fig.~\ref{fig:npf},c-d. System-size analysis shows that the peak culling time grows algebraically with system size, ${\rm max}(\langle M\rangle)\propto L^{\alpha}$, see fig.~\ref{fig:exponents}a. The exponents we find are $\alpha=1.22$ in 2D that compares well with a previous study~\cite{JengSchwarz}, and $\alpha=1.33$ in 3D.

We probe a diverging \emph{length-scale} at the transition by finite-size scaling. Fig.~\ref{fig:npf},a-b shows how the transition becomes sharper as the system size is increased. In the thermodynamic limit, we expect the typical cluster-size to diverge as $\exp\left[A \left(\rho_c-\rho\right)^{-\mu}\right]$, where $\mu=\nu(1-1/z)$. The values of the DP exponents are $\nu=1.74(2D),1.29(3D)$ and $z=1.58(2D),1.77(3D)$~\cite{DP_exponents}, thus we expect $\mu=0.63(2D),0.56(3D)$. A system of linear dimension $L$ jams at density $\rho_c(L)$ for which this cluster size equals the system size: $\exp\left[A\left(\rho_c(\infty)-\rho_c(L)\right)^{-\mu}\right]\propto L$, where $\rho_c(\infty)$ is the theoretical value which we expect to obtain in the infinite system-size limit. Inferring $\mu$ from $\rho_c(L)$ is impractical because even for the largest systems we can simulate ($L^3 \approx 0.5 \cdot 10^9$) $\rho_c(L) \approx 0.35$ is too far from the 3D DP value of $\rho_c(\infty)\approx 0.43$. Instead of looking at the system-size dependence of $\rho_c$, we claim~\cite{JengSchwarz} that the transition width $W$ is proportional to $\rho_c(\infty)-\rho_c(L)$, and therefore expect to have $W\propto(\ln L)^{-1/\mu}$. This functional form agrees with our numerical results, see fig.~\ref{fig:exponents}b, but with scaling exponents $\mu=0.38(2D),0.27(3D)$ which are about a factor of two smaller from the theoretical values given above. This discrepancy has been previously observed for the 2D spiral model and attributed to the limited range of systems sizes probed in such simulations~\cite{JengSchwarz}.

\section{Dynamics} 

The static properties of our 3D model are qualitatively similar to those of the previously studied 2D spiral model. Interestingly, the dynamics of these two models are qualitatively different. In both models, the backbone of frozen particles is comprised of perpendicular sets of DP chains. In the spiral model, these chains form a 2D mesh, and mobile particles are confined within the cells of this mesh (see fig.~\ref{fig:face}a). Thus, above the critical density $\rho_c^{2D}$, for which these percolating clusters of permanently-frozen particles appear, the mobile particles can diffuse only over finite distances, and the mean-square-displacement saturates with time. In our 3D model, this mesh is also comprised of 1D chains of permanently-frozen particles, hence the mobile particles can use the third dimension to travel between the cells of this mesh (see fig.~\ref{fig:face}b) and the long-time behavior may be diffusive. 

Fig.~\ref{fig:npf},e-f shows that in 3D diffusivity is indeed finite also above $\rho_c^{3D}$, whereas in 2D it vanishes at $\rho_c^{2D}$. Interestingly, the diffusion coefficient does not seem to exhibit any singularity at $\rho_c^{3D}$. This is because below $\rho_c^{3D}$ although all particles eventually move, many of them are already blocked by a hierarchy of many other particles (as may be seen from the huge values of the culling time there), and the particle-averaged mean-square-displacement is dominated by the particles that do manage to move more quickly (and do not require many culling iterations in order to be removed in the analysis of which particles are frozen). For the diffusion of these particles, no qualitative change occurs at $\rho_c^{3D}$, thus $D$ is smooth there. In our 3D model, above $\rho_c^{3D}$ diffusing particles travel through windows in the mesh of string-like DP chains. As the density is increased beyond $\rho_c^{3D}$, more particles are permanently frozen, thus these chains become thicker, until at some second critical density $\rho_D^{3D}$ the windows in the mesh close, such that particles can no longer diffuse between neighboring cages, and the diffusivity should vanish. $\rho_D^{3D}$ is the critical density for directed percolation of a surface in 3D, which should be bounded from above by the critical density $\rho_s^{3D}=0.69$ of surface percolation in 3D~\cite{Aizenman1983}.

\section{Conclusions}

In summary, we propose the first 3D jamming-percolation model, and provide a theoretical proof that it exhibits a mixed-order jamming transition at a non-trivial critical density. The system-size dependence of the transition width indicates that the length scale associated with the size of frozen clusters diverges faster than a power law. The 3D divergent length-scale exponent is slightly smaller than in 2D. Many theoretical studies, including those based on lattice models, as well as recent experiments done on glassy or jammed matter use 2D or quasi-2D systems~\cite{Han,Ghosh,Chen,Bi}. The behavior in lower dimensions may differ from that of the corresponding 3D systems. A relevant question thus remains whether substantial differences occur when going from two to three dimensions and what would be the origin of such differences. The dynamic results we present here are one example showing that diffusive behavior in 3D differs from that of 2D systems. This occurs due to the way percolating structures span 3D systems, providing an additional dimension for the particles to diffuse in. This dimensionality dependence, related to diffusion in porous media and in chemical gels, is quite general. It would be interesting to test to what extent this is related to the well-known decoupling of self diffusion and viscosity in glasses. It is certainly useful that we have a concrete glassy example where the diffusion coefficient is still non-zero but the structure does not relax anymore. Thus, our simple kinetically-constrained lattice model provides scope for future studies of various phenomena related to glassiness, jamming and dynamics in three dimensions.

\acknowledgments

We thank Haye Hinrichsen, Dov Levine, and Andrea Liu for helpful discussions, and Amir Natan for computing resources. This research was supported by the Israel Science Foundation grants No. $617/12$, $1730/12$.

\renewcommand{\theequation}{A\arabic{equation}}
\setcounter{equation}{0}

\section{Appendix A - Proof for Exponential Bound on Path Probabilities}

We want to prove that in 3D DP, for $\rho<\rho_c^{3D}$ the probability that a site is part of a directed path of $n$ sites is smaller than or equal to $e^{-n/\xi}$, where $\xi$ is finite and depends on the density. Let ${\cal B}_{n}=\left[0,n\right]^{3}$ be a cube of side $n+1$, and $\partial{\cal B}_n={\cal B}_{n}\setminus{\cal B}_{n-1}$ its boundary. We denote the number of sites on the boundary $\partial{\cal B}_{n}$ which have a directed path to them starting from the origin $\vec{0}=(0,0,0)$ by $N_{n}$. The total number of sites in an infinite lattice that have a directed path to them from the origin is thus $N_{tot}=\sum^{\infty}_{n=0}E(N_{n})$, where $E(N_{n})$ is the expectation value of $N_{n}$. Since $\rho<\rho_c^{3D}$, the size $N_{tot}$ of the DP cluster starting from the origin is finite. Hence, there must be some $m$ for which $E(N_{m})\leq\frac{1}{2}$.

Consider the probability $P(\vec{0}\rightarrow\partial{\cal B}_{m+k})$ that the origin has a directed path to the boundary of the cube ${\cal B}_{m+k}$ with some $k\geq0$, i.e. that the origin has a directed path of length $m+k+1$ starting from it. This is equal to the probability that there is a directed path from the origin to some site $\vec{r}$ on the boundary of a smaller cube $\partial{\cal B}_{m}$ and from that site to the boundary of the larger cube
\begin{eqnarray}
& & P(\vec{0}\rightarrow\partial{\cal B}_{m+k})= \nonumber\\ & & =P\left(\exists \vec{r} \in\partial{\cal B}_{m},\left\{\vec{0}\rightarrow \vec{r}\right\}\wedge\left\{\vec{r}\rightarrow\partial{\cal B}_{m+k}\right\}\right) .  
\end{eqnarray}
The probability that the site $\vec{r}$ is connected to the larger boundary is smaller or equal to the probability that it is connected to the boundary of a cube with one corner at $\vec{r}$ and the opposite corner at $\vec{r}+(k,k,k)$, because in order to reach the boundary $\partial{\cal B}_{m+k}$ it must first pass through the boundary of $\vec{r}+\partial{\cal B}_{k}$. Hence,
\begin{eqnarray}
& & P(\vec{0}\rightarrow\partial{\cal B}_{m+k})\leq\nonumber\\
& & \leq P\left(\exists \vec{r}\in\partial{\cal B}_{m},\left\{\vec{0}\rightarrow \vec{r}\right\}\wedge\left\{\vec{r}\rightarrow \vec{r}+\partial{\cal B}_{k}\right\}\right)\nonumber\\
& & \leq\sum_{\vec{r}\in\partial{\cal B}_{m}}P\left(\left\{\vec{0}\rightarrow \vec{r}\right\}\wedge\left\{\vec{r}\rightarrow \vec{r}+\partial{\cal B}_{k}\right\}\right) .
\end{eqnarray}
The probabilities that the origin is connected to $\vec{r}$ and that $\vec{r}$ is connected to the boundary are independent, and so 
\begin{eqnarray}
& & P(\vec{0}\rightarrow\partial{\cal B}_{m+k})\leq\nonumber\\
& & \leq\sum_{\vec{r}\in\partial{\cal B}_{m}}P\left(\vec{0}\rightarrow \vec{r}\right)P\left(\vec{r}\rightarrow \vec{r}+\partial{\cal B}_{k}\right)\nonumber\\
& & =\sum_{\vec{r}\in\partial{\cal B}_{m}}P\left(\vec{0}\rightarrow \vec{r}\right)P\left(\vec{0}\rightarrow\partial{\cal B}_{k}\right)\nonumber\\
& & =E(N_{m})P\left(\vec{0}\rightarrow\partial{\cal B}_{k}\right)\leq\frac{1}{2}P\left(\vec{0}\rightarrow\partial{\cal B}_{k}\right) ,
\end{eqnarray}
since we can choose $m$ such that $E(N_{m})\leq\frac{1}{2}$. For any $n>m$, we write $n=mg+s$ where $g$ and $s$ are integers with $0\leq s \leq m$. Thus $P(\vec{0}\rightarrow\partial{\cal B}_{n})\leq\left(\frac{1}{2}\right)^{g}\leq\left(\frac{1}{2}\right)^{n/m-1}$, which decays exponentially with $n$, as required.

\section{Appendix B - Calculation of Bound in Eq.~(\ref{eq:one})}

We want to show that
\begin{eqnarray}
\tilde{P} &\equiv& 18\prod^{\ell}_{s=1}\left(1-e^{-s/\xi}\right)\left(1-e^{-\left(\ell-s+1\right)/\xi}\right)\left(1-e^{-\ell/\xi}\right)\nonumber\\ &\geq& C \exp\left(-\ell e^{-\ell/\xi}\right) .
\end{eqnarray}
We first note that by symmetry $\prod^{\ell}_{s=1}\left(1-e^{-s/\xi}\right)= \prod^{\ell}_{s=1}\left(1-e^{-\left(\ell-s+1\right)/\xi}\right)$. Such that
\begin{eqnarray}
\tilde{P} &=& 18\left(1-e^{-\ell/\xi}\right)^{\ell}\prod^{\ell}_{s=1}\left(1-e^{-s/\xi}\right)^2 \nonumber\\ &=& 18\left(1-e^{-\ell/\xi}\right)^{\ell}\exp\left[2\sum^{\ell}_{s=1}\ln\left(1-e^{-s/\xi}\right)\right] .
\end{eqnarray}
The argument of the logarithm is smaller than $1$ for all $s\geq1$, and so adding more terms to the sum would only decrease it. Hence we can write $\tilde{P}\geq18\left(1-e^{-\ell/\xi}\right)^{\ell}\exp\left[2\sum^{\infty}_{s=0}\ln\left(1-e^{-s/\xi}\right)\right] $. In order to show that the infinite sum converges to a finite value, we change it to an integral over $w=s/\xi$, $\sum^{\infty}_{s=0}\ln\left(1-e^{-s/\xi}\right)\approx\xi\int^{\infty}_{0}\ln\left(1-e^{-w}\right)dw=-\frac{\pi^{2}\xi}{6}$, and thus $\tilde{P} \geq C\left(1-e^{-\ell/\xi}\right)^{\ell}$, with $C= 18\exp\left(-\frac{\pi^{2}\xi}{3}\right) > 0$. For finite $\xi$ and large $\ell$, we may approximate
\begin{eqnarray}
\left(1-e^{-\ell/\xi}\right)^{\ell} &=& \left[\left(1-e^{-\ell/\xi}\right)^{e^{\ell/\xi}}\right]^{\ell e^{-\ell/\xi}} \nonumber\\ &\approx& \exp\left(-\ell e^{-\ell/\xi}\right) .
\end{eqnarray}


\end{document}